# XR is XR:
# Rethinking MR and XR as Neutral Umbrella Terms

Takeshi Kurata *

*Research Institute on Human and Societal Augmentation, AIST

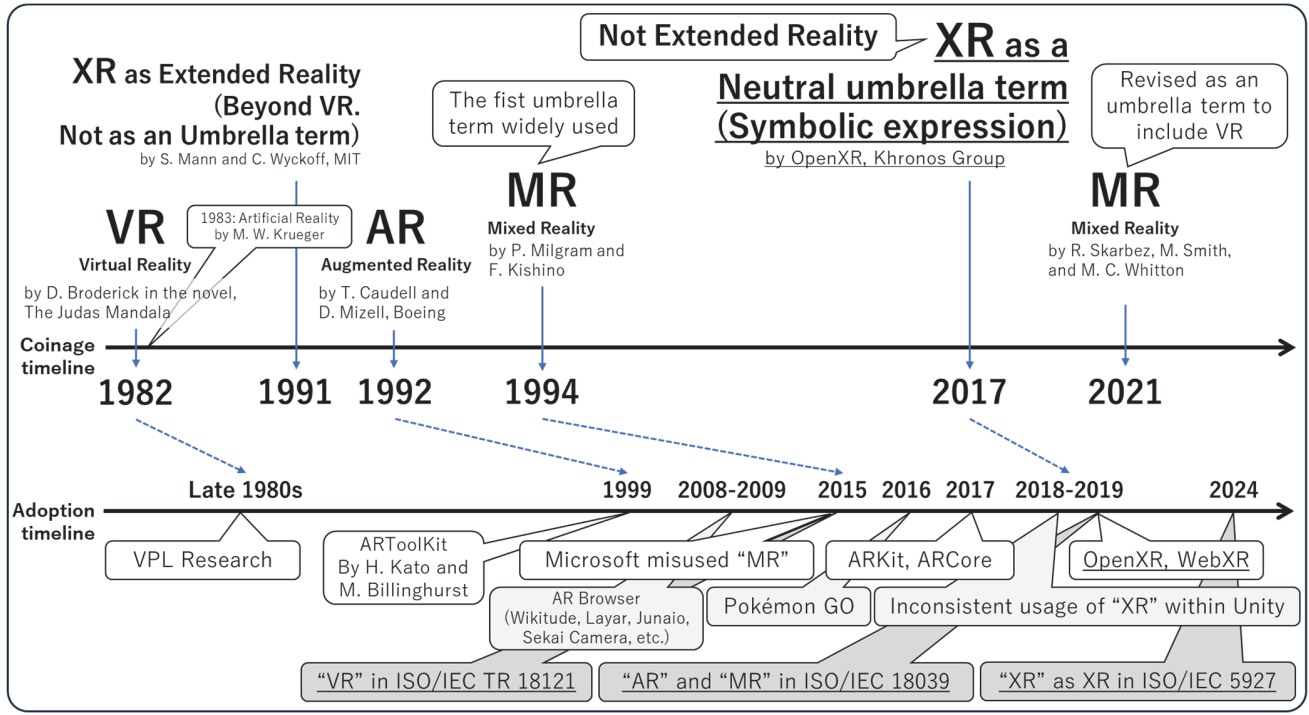

Figure 1: Timeline of the emergence and evolution of XR-related terminology. The timeline highlights key milestones, including VR (Virtual Reality) (1982), AR (Augmented Reality) (1992), and MR (Mixed Reality) (1994), followed by the introduction of XR as a neutral and inclusive label in standardization contexts (e.g., OpenXR), as well as subsequent reinterpretations of MR. The figure distinguishes between the coinage timeline, indicating when key terms were introduced, and the adoption timeline, which captures key drivers and events that contributed to the adoption of these terms in industry and society. The coinage timeline highlights key milestones, including VR (Virtual Reality) (1982), AR (Augmented Reality) (1992), and MR (Mixed Reality) (1994), followed by the introduction of XR as a neutral and inclusive label in standardization contexts (e.g., OpenXR), as well as subsequent reinterpretations of MR. In contrast, the adoption timeline reflects the spread of these terms through technological developments, platforms, and applications, illustrating how their usage has expanded and, in some cases, diverged from their original definitions. Items related to standardization activities (e.g., Khronos and ISO/IEC) are underlined to emphasize their role in shaping and stabilizing terminology across different stakeholders.

**ABSTRACT**

The term XR is currently widely used as an expression encompassing Virtual Reality (VR), Augmented Reality (AR), and Mixed Reality (MR). However, there is no clear consensus regarding its origin or meaning. XR is sometimes explained as an abbreviation for Extended Reality, but multiple interpretations exist regarding its etymology and formation process. This paper organizes the historical formation of terminology related to VR, AR, MR, and XR, and reexamines the context in which the term XR emerged and how it has spread. In particular, by presenting a timeline that distinguishes between the coinage of terms and the drivers of their adoption, we suggest that XR, as an umbrella term, functions not as an abbreviation of *Extended Reality*, but rather as a neutral symbolic label that encompasses multiple "reality"-related terms. Furthermore, we argue that stable usage of terminology, including XR, requires governance through collaboration among academia, industry, and standardization organizations.

**Index terms**: XR, MR, AR, VR, Terminology, Taxonomy, Reality–Virtuality Continuum, Semantic Interoperability, Standardization

## 1 INTRODUCTION

For concepts that span multiple domains, such as XR and the metaverse, ambiguity in terminology can hinder communication among research communities, industry, and policy sectors. In software engineering, APIs and data formats are standardized to enable interoperability between different systems. Similarly, terminology and its underlying semantic structure can be understood as a form of "semantic interoperability" that enables coordination among diverse stakeholders [1].

However, there is no clear consensus even within the research community regarding the origin or meaning of the term XR. As a result, fundamental questions remain unresolved, such as whether XR should be interpreted as an abbreviation of Extended Reality or understood as a neutral symbolic label. These issues were discussed at the ISMAR 2024 panel session "xR is dead, long live XR!" [2], and although a standardization committee was established at ISMAR 2025 [3] and the XRStand 2025 workshop [4] was also held, to continue discussions, it was confirmed that no common understanding exists within the research community regarding the origin or definition of XR.

This paper first organizes the historical background of XR-related terminology and examines the contexts in which XR emerged. Based on this analysis, we argue that XR should be understood not as an abbreviation of a specific term but as a neutral and inclusive symbolic label. In this context, we also consider the role of MR, which functioned as an umbrella term prior to the widespread adoption of XR and has recently been subject to reinterpretation. Furthermore, we discuss how technical terminology has been formed through interactions among academia, industry, and standardization bodies, and address future challenges in standardization, including the need for stronger engagement of domain experts.

## 2 HISTORY OF XR-RELATED TERMINOLOGY

This section organizes the emergence of major terms related to VR, AR, MR, and XR, as well as changes in their meanings. The focus here is not the research history of XR-related technologies themselves, but rather the history of terminology, namely how these terms were introduced and interpreted in different contexts.

As shown in Figure 1, the history of XR-related terminology can be understood from two complementary perspectives: the coinage timeline, which represents when key terms were introduced, and the adoption timeline, which captures key events and technological developments that contributed to the adoption of these terms in industry and society. This distinction is important because the introduction of a term and its subsequent adoption are often separated in time and influenced by different factors.

The term Virtual Reality (VR) gained prominence in the late 1980s, particularly through the activities of VPL Research [5][6]. The term is known to have appeared as early as in Damien Broderick's novel The Judas Mandala [7]. VR generally refers to technologies that immerse users in computer-generated virtual environments. Although the concept of VR existed earlier, and related terms such as Artificial Reality had already been proposed [8], the term "Virtual Reality" gained widespread attention in the late 1980s.

An early example of the expression XR is Extended Reality, proposed in 1991 by Steve Mann and Charles Wyckoff [9]. In this context, Extended Reality referred to environments that extend human perception beyond VR; it was not proposed as an umbrella term encompassing VR, AR, and MR. At that time, AR (1992) and MR (1994) had not yet been proposed, and therefore Extended Reality was not intended as an umbrella term encompassing multiple "reality" concepts.

The term Augmented Reality (AR) was proposed in 1992 by Tom Caudell and David Mizell [10]. AR refers to technology that overlays virtual information onto the real world. Subsequently, Azuma [11] clarified its definition through three characteristics: the integration of real and virtual environments, real-time interaction, and three-dimensional registration. The adoption timeline in Figure 1 shows that the widespread use of AR was driven largely by technological platforms, applications, and consumer-facing services, such as ARToolKit [12], AR browsers [13][14][15][16], mobile AR systems [17][18], and applications like Pokémon GO [19].

Milgram and Kishino proposed the Reality-Virtuality Continuum in 1994 [20], positioning MR as an intermediate domain between real and virtual environments. As a result, MR functioned as one of the earliest umbrella terms for reality-related technologies, including Augmented Virtuality (AV).

In parallel with these developments, international standardization efforts have also formalized definitions of related concepts. For example, ISO/IEC defines VR, AR, and MR within its standard frameworks [21][22], providing a structured basis for understanding these terms across different domains.

## 3 EMERGENCE OF XR AS A NEUTRAL SYMBOL

As discussed in Section 2, MR functioned as an umbrella term for reality-related technologies. However, its conceptual scope was not always sufficient for consistently describing the relationships among VR, AR, and related domains. In Milgram's definition, MR is positioned as an intermediate domain between the real and the virtual, and can be interpreted as not including fully virtual VR. Consequently, when referring precisely to both VR and MR, it became necessary to explicitly use expressions such as "VR and MR." This reduced the convenience of MR as an umbrella term and may have contributed to the increasing complexity of terminology when referring collectively to VR, AR, and MR.

It may also be noted that the title of the international conference ISMAR is somewhat unusual. However, this resulted from merging two predecessor conferences (IWAR/ISAR and ISMR) while respecting both, and thus differs in intent from the discussion of umbrella terminology.

Following the introduction of MR, Microsoft announced HoloLens in 2015 [23][24] and used the term "Mixed Reality" to describe AR experiences based on optical see-through displays. While not entirely incorrect, this usage referred to only a subset of the domain that MR was intended to cover, contributing to inconsistencies between conceptual definitions and industrial usage.

In this confusion of terminology, the Khronos Group initiated the development of an open standard API, later named OpenXR, to enable interoperability across VR, AR, and related systems [25][26]. Within this effort, a key policy was adopted that terminology should remain neutral and independent of specific interpretations of existing reality-related terms. Under this policy, the symbol "XR" was introduced, representing "any reality you like." In this context, the "X" does not denote a specific word such as "Extended" or "Cross," but instead functions as a symbolic placeholder that emphasizes neutrality and inclusiveness [1][26].

A similar interpretation can also be found in web-based standardization efforts. WebXR [27], developed within the W3C Immersive Web Community Group and standardized by the Immersive Web Working Group, describes "XR" not as a fixed abbreviation but as a flexible placeholder, for example "Your Reality Here," that encompasses various forms of reality, including VR and AR. As indicated in Figure 1, such standardization efforts also contributed to the adoption of XR terminology. This usage reinforces the view that XR functions as a neutral symbolic label rather than a term derived from a specific expansion such as Extended Reality.

Such design approaches reflect the role of Standards Development Organizations (SDOs) in providing terminology that can be consistently used across stakeholders. In contrast, such a position is generally difficult for entities acting as terminology promoters, which tend to emphasize differentiation, as illustrated in Figure 2.

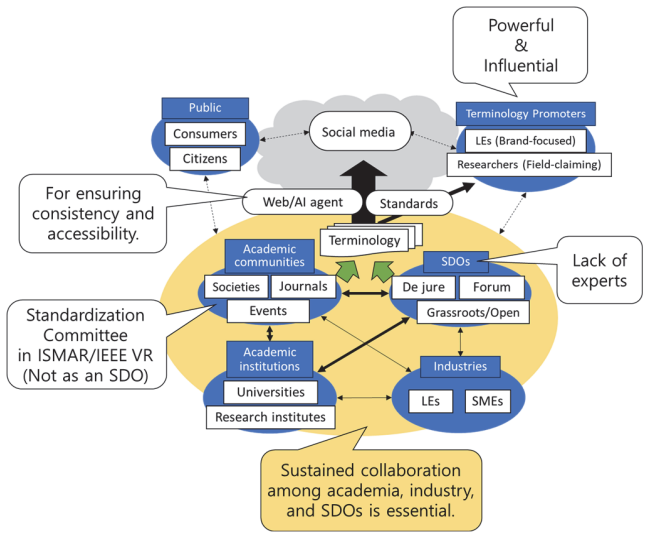

Figure 2: Stakeholders involved in the formation, dissemination, and maintenance of terminology across research, standardization, industry, and the public. Revised from Figure. 2 in [1].

## 4 SPREAD OF XR AS AN UMBRELLA TERM AND REINTERPRETATION OF MR

In recent years, XR has become widely used as a comprehensive expression encompassing various reality-related terms such as VR, AR, and MR. As indicated in the adoption timeline of Figure 1, this spread is closely associated with technological developments, platform ecosystems, and standardization efforts such as OpenXR and WebXR, which facilitated interoperability and accessibility.

On the other hand, since the term Extended Reality already existed, XR is sometimes interpreted as its abbreviation. While XR can indeed stand for Extended Reality, the discussion above and Figure 1 suggest that XR, as an umbrella term, is more appropriately understood not as an abbreviation of any specific term, but as a neutral symbol in its own right. Furthermore, there is little evidence that XR was used as an umbrella term prior to around 2017, while MR largely fulfilled that role. This suggests that XR, as an umbrella term, is best understood based on definitions established by SDOs such as the Khronos Group and W3C.

In 2018, Mann and colleagues reused the term Extended Reality and further proposed Multimediated Reality as a higher-level concept [28]. Although the present paper focuses on terminology rather than concepts, if XR is regarded as an umbrella term for various realities such as VR and AR, then Multimediated Reality should also be positioned within XR. Conversely, since Multimediated Reality is a superordinate concept of Extended Reality, interpreting XR as an abbreviation of Extended Reality introduces a conceptual inconsistency.

In 2021, Skarbez et al. revisited the Reality-Virtuality Continuum and proposed a framework that reinterprets the scope of MR [29][30]. In this framework, MR may be understood as including forms of VR that are currently realizable, while fully synthetic or purely virtual environments may remain outside its scope, suggesting the possibility of reusing MR as an umbrella term. At the same time, it highlights how terminology may evolve through reinterpretation while maintaining continuity with earlier conceptual frameworks. Based on this reinterpretation, if an umbrella term is required within the Reality-Virtuality continuum, it is not necessarily limited to XR, and MR may also serve that role.

## 5 CONCLUSION

This paper organized the origins and formation processes of XR-related terminology by distinguishing between the coinage of terms and the drivers of their adoption. The analysis shows that terminology in this domain remains unstable due to multiple factors, including differing interpretations of MR, marketing-driven usage, and competing interpretations of XR.

Technical terminology, including XR, is formed through interactions among academia, industry, and SDOs. While academia and industry often emphasize differentiation, standardization requires neutrality, reflecting differences in roles and incentives among stakeholders. XR as a symbolic label can be understood as emerging from such requirements for neutrality in standardization contexts. Similarly, ISO/IEC also defines XR as a symbolic umbrella term [31].

Historically, terminology in this field often emerged through proposals by individual researchers or organizations and subsequently spread through adoption. However, recent developments indicate a shift toward terminology being established through consensus among multiple stakeholders within SDOs. This transition reflects an increasing need for shared and interoperable terminology across diverse domains.

Figure 2 is a revised version of the figure presented in [1], illustrating the main stakeholders involved in the formation, dissemination, and maintenance of terminology and their relationships. Understanding the history of terminology and the governance among related stakeholders, and appropriately recognizing each party's position and the role of terminology, is essential for improving semantic interoperability in research and technological development, as well as for ensuring fair and effective innovation and dissemination of technology. Such structures and governance are not limited to terminology, but are also broadly relevant to standardization processes in general, particularly in emerging domains such as XR.

Inconsistencies in terminology are not limited to differences across organizations, but can also be observed within individual actors. For example, Unity Technologies presents differing interpretations of XR across its official documentation and glossary [32][33]. Although Unity does not necessarily act as a terminology promoter in the strict sense, its significant influence in the ecosystem suggests that such inconsistencies may contribute to the current ambiguity in terminology. This example illustrates how inconsistencies can arise even in influential actors, reinforcing the need for coordinated governance of terminology across stakeholders.

In this context, SDOs are expected to play a central role in developing and maintaining standards while connecting stakeholders. However, they suffer from a chronic shortage of experts. In practice, there are cases where individuals with limited expertise are responsible for the development and review of international standards. Addressing this issue and fostering broader engagement from both academia and industry in standardization activities remain important challenges. For example, in addition to ISMAR 2025, a standardization committee has also been established in IEEE VR 2026 [34], and panel sessions have been organized to promote discussion and participation [35]. Through such efforts, it is expected that more experts from both academia and industry will become engaged in standardization activities over time.

### ACKNOWLEDGEMENT

I would like to thank the members of the standardization committees of IEEE VR and ISMAR for their contributions to standardization activities, including Jean Botev (University of Luxembourg, Luxembourg), Dooyoung Kim (KAIST, South

Korea), Seonji Kim (KAIST, South Korea), Seoyoung Kang (KAIST, South Korea), Ryosuke Ichikari (AIST, Japan), Yahya (Yohan) Hmaiti (University of Central Florida, USA), Jen-Shuo Liu (Columbia University, USA), Mykola Maslych (University of Central Florida, USA), Trond Nilsen (University of Washington, USA), Christine Perey (Spime Wrangler, Switzerland), and Abolghasem Sadeghi-Niaraki (Sejong University, South Korea).

I also thank the panelists, moderators, and speakers of related events for their valuable insights shared through discussions and presentations, including Mark Billinghurst (University of South Australia), Wolfgang Broll (TU Ilmenau), Takefumi Hiraki (Cluster, Inc. / University of Tsukuba, Japan), Kiyoshi Kiyokawa (Nara Institute of Science and Technology), Missie Smith (Auburn University), Richard Skarbez (La Trobe University, Australia), J. Edward Swan II (Mississippi State University, USA), Neil Trevett (NVIDIA, USA), and Tim Weissker (RWTH Aachen University, Germany).